# Optical vortex produced by Single Photon Subtraction from two mode squeezed state Produces Maximum Entanglement

Abir Bandyopadhyay[1,2], Anindya Banerji[1] and R. P. Singh[2]
[1]Hooghly Engineering & Technology College, Pipulpati, Hooghly 712103, INDIA
[2]Physical Research Laboratory, Navrangpura, Ahmedabad 380009, INDIA

**Abstract**
In this article we report that the entanglement produced by single photon subtraction is maximum, by studying the entanglement of multi photon subtraction from two mode squeezed states. We argue that the single photon subtraction produces maximum entanglement as it is more non-classical in nature than many photon subtractions.
**OCIS:** 270.5585, 060.5565, 190.4180, 200.3050

## 1. Introduction

Optical vortices or phase singularities in the field of light, described by a non-separable two-dimensional field, have drawn a great deal of attention in the last two decades [1] and these have prompted to start a new branch in physical optics known as singular optics [2]. Circular optical vortex beams with helical wave front can be produced in a controlled manner using methods such as computer-generated hologram (CGH), cylindrical lens mode converter, and spiral phase plate [1]. Because of their specific spatial structure and associated orbital angular momentum (OAM), they find profound applications in the field of optical manipulation [3], optical communication [4], quantum information and computation [5]. For a vortex, the OAM follows the same numbers as vorticity or topological charge of the vortex i.e. an optical vortex of topological charge *m*, carries an OAM of *mℏ* per photon. This has another significant implication. It implies that the OAM associated with an optical vortex is also quantized, though this aspect has hardly been studied [6–10]. Most applications of vortices formed by quantized radiation fields lie in the classical realm. It has been shown that such states can be generated from two mode squeezed vacuum under a linear transformation belonging to the SU(2) group with some restrictions.

We studied the Wigner function of a general elliptical vortex, the circular vortex being a special case of an elliptical vortex [8]. At this point, we must make a distinction between the optical vortices of [1] and the vortex states discussed in [6-9] as well as considered in the present article. The optical vortices considered in [1] exist in real space of the classical wave field, while the vortex states considered in [6-9] and in this article are quantum mechanical wave packets formed by two quantized radiation modes and exist in the abstract space of the modes.

Recently Agarwal has proposed quantum optical vortex by subtracting a photon by a 99% transmitting beam-splitter (BS) from one of the two modes of spontaneous parametric down-conversion (SPDC), generated by the non-linear crystal (NLC) as shown in Fig. (1). He showed that the subtraction from one mode results into an addition to the other mode; it produces a vortex, which is elliptical in shape. It must be noted that a single photon subtraction or addition is not just a theoretical construct rather it has already demonstrated by V. Parighi et al. [11]. A more recent paper studied the enhancement of entanglement, though defined differently, in photon subtraction/addition [12].

The paper is organized as follows. In the second section we first describe the experimental proposal of Agarwal for completeness of this paper. We comment that the general vortex of [9] state has a spatial distribution, which matches our previous work [8]. We extend the experimental procedure and calculate the entanglement ratio with the corresponding state with two mode squeezed state. We plot them to reproduce the single photon subtraction results of [9]. We also compute and plot the entanglement for multi-photon subtraction and find that vortex of unit vorticity produces optimum entanglement. We argue that the reason of such behavior is due to the high non-classicality of single photon. Finally we conclude our findings.

## 2. Entanglement of vortex states produced by Photon Subtraction from the output of SPDC

We now consider the engineering of a vortex state in quadrature space. Let us consider the arrangement shown in Fig. 1. Here we use a down converter to produce the two–mode squeezed state $|\xi\rangle$ of the signal (*s*) and the idler (*i*) modes *a* and *b*. The two mode squeezed field (output from SPDC) is defined as

$$|\xi\rangle = \exp(\xi a^\dagger b^\dagger - \xi^* ab)|0,0\rangle, \quad \xi = re^{i\theta}, \qquad (1)$$

with $\xi$ as a complex parameter. *a* and *b* are the usual mode operators satisfying the general commutation relations: $[a, a^\dagger] = [b, b^\dagger] = 1$, $[a, b^\dagger] = 0$, etc.

Now one photon can be subtracted from mode *b* with the help of a beam splitter and that can be detected by a single photon detector. The probability of detecting is proportional to the square of reflectivity. The resulting state $|\xi\rangle^{(s)}$ can be obtained by using the suitable transformations as

$$|\xi\rangle^{(s)} = \mathcal{B}\exp(\xi a^\dagger b^\dagger - \xi^* ab)(b\cosh r + a^\dagger \sinh r e^{i\theta})|0,0\rangle \qquad (4)$$

Where $\mathcal{B}$ is the normalization constant. From the above equation one can obtain the following form

$$|\xi\rangle^{(s)} = \exp(\xi a^\dagger b^\dagger - \xi^* ab)|1,0\rangle e^{i\theta}. \qquad (5)$$

This state is obtained by applying the squeezing transformation to a state with a single photon in mode $a$. This state exhibits many nonclassical properties due to the presence of the Fock state as well as the squeezing operator. One can also write the state (5) in terms of Fock states by using the decomposition of the squeezing operator and obtain

$$|\xi\rangle^{(s)} = \frac{e^{i\theta}}{\cosh^2 r}\sum_{m=0}^{\infty} e^{im\theta}(\tanh r)^m \sqrt{n+1}|n+1,n\rangle = e^{i\theta} a^\dagger \frac{|\xi\rangle}{\cosh r}. \quad (6)$$

We see that the state (6) can also be obtained by adding a photon to $|\xi\rangle$ in the mode $a$, which is similar to Parighi et al.'s experiment. The state (6) exhibits the property that there is a difference of one photon between the signal and the idler states. In literature, this is one of the so called "pair coherent states", where there is a fixed difference in the number of photons in the two modes [13, 14]. Pair coherent states provide an interesting example of non-classical states of the two-mode radiation field [13]. It has been studied in detail for their non-classical properties and as examples of EPR states [13]. More recently, their experimental signatures have also been explored [15].

The other property of interest is the vortex structure. One can calculate the quadrature distribution for the signal state:

$$\Psi^{(s)}(x_a, x_b) = \frac{\sqrt{2}e^{i\theta}(x_a - \kappa x_b)}{(1-\kappa^2)^{3/2}\pi^{1/2}\cosh^2 r} \exp\left[\frac{2x_a x_b \kappa - (x_a^2 x_b^2)\kappa^2}{1-\kappa^2} - \frac{1}{2}(x_a^2 + x_b^2)\right].$$

where quadrature operators $x_a, x_b, y_a, y_b$ are defined as $a = \frac{x_a + iy_a}{\sqrt{2}}$, $b = \frac{x_b + iy_b}{\sqrt{2}}$ with $\kappa = \tanh r\, e^{i\theta}$. If $\kappa = i|\kappa|$ then $\Psi^{(s)}(x_a, x_b)$ shows vortex structure. The $|\Psi^{(s)}(x_a, x_b)|^2$ shows a clear vortex structure and the elliptic nature is also evident as $|\kappa| \neq 1$ [8, 9].

The log negativity parameter defined by

$$\varepsilon = \log_2(1 + 2\mathbb{N}), \quad (11)$$

is used to quantitatively study the entanglement in the vortex state, where $\mathbb{N}$ is the modulus of the sum of all the negative eigenvalues associated with the partial transpose $\rho^{\text{PT}}$ of the density matrix $|\xi\rangle^{(s)(s)}\langle\xi|$. From equation (6), the partial transpose is

$$\rho^{\text{PT}} = \sum_{m,n} c_m c_n |m+1, n\rangle\langle n+1, m| e^{i(m-n)\theta}, \quad (12)$$

$$c_p = (\tanh r)^p \frac{\sqrt{p+1}}{\cosh^2 r}. \quad (13)$$

where $p = m, n$. Therefore the log negativity parameter (11) becomes

$$\varepsilon = \log_2\left(\sum_n c_n\right)^2. \quad (14)$$

This sum can be computed numerically. A similar calculation for the two mode squeezed vacuum gives the result $\log_2(e^{2r})$ using $c_n=(\tanh r)^n/\cosh r$. The entanglement in the vortex state can be compared with that in the state $|\xi\rangle$ by studying the ratio of log negativities of the two states

$$\tilde{\varepsilon} = \left(\sum_n c_n e^{-r}\right)^2. \tag{16}$$

Since $\tilde{\varepsilon} > 1$, it can be concluded that in a sense the vortex state $|\xi\rangle^{(s)}$ is more entangled than the two – mode squeezed vacuum state $|\xi\rangle$. This plot is reproduced with our code in Fig. (2a).

We next consider the case of multiphoton subtraction from one of the modes of SPDC, which is a two mode squeezed state. The multiphoton subtraction can be obtained by two ways. Firstly, if the reflectivity of the BS is increased to 2%, there will be a four photon subtraction in the output for a $10^4$ photon intense beam. In the other way, we may repeat the single photon subtraction multiple times to generate two/three photon subtraction.

We compute the same ratio $\tilde{\varepsilon}$ defining the relative measure of entanglement for the multiphoton subtracted two mode squeezed state and plot them in Figs. (2b-d). First observation is that there is a very small increase for two photon subtraction which falls sharply after reaching the peak. Moreover, further subtraction worsens the situation. For three photon subtraction the initial value falls to half as shown in Fig. 2(c). For four photon subtraction, the starting value is 0.042 as shown in Fig. 2(d). We didn't show in the Fig. (2), but want to mention that for five photon subtraction it starts from a value of the order of $10^{-6}$. Though such a drastic behavior cannot be explained physically, we argue that single photon state is more non-classical in nature than all the set of Fock states. Thus the quantum nature of entanglement is more pronounced in the case of single photon subtraction.

## 3. Conclusion:

To conclude, we have studied multiphoton subtraction in this article. We reproduced Agarwal's result of single photon subtraction as part of our work. We have shown that single photon subtraction produced maximum entanglement. We argue that subtraction of higher number of photons reduces the non-classical nature of the state.

## 4. Acknowledgement:

This work has been supported by DST through SERB grant no.: SR/S2/CMP-0002/2011. A. Bandyopadhyay acknowledges PRL for Associateship, where part of the work has been done, and G. S. Agarwal for useful discussion.

# Figures

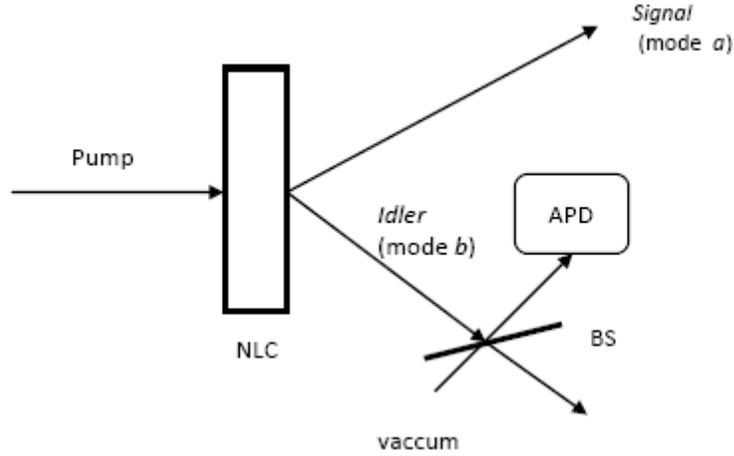

**Figure 1**: Method for production of vortex state by photon subtraction from a two – mode squeezed states using a beam splitter. Here NLC is a nonlinear crystal. SPDC, the signal mode (*a*) and the idler mode (*b*) are shown. APD is Avalanche Photo Diode and BS is a beam splitter [9].

(a)
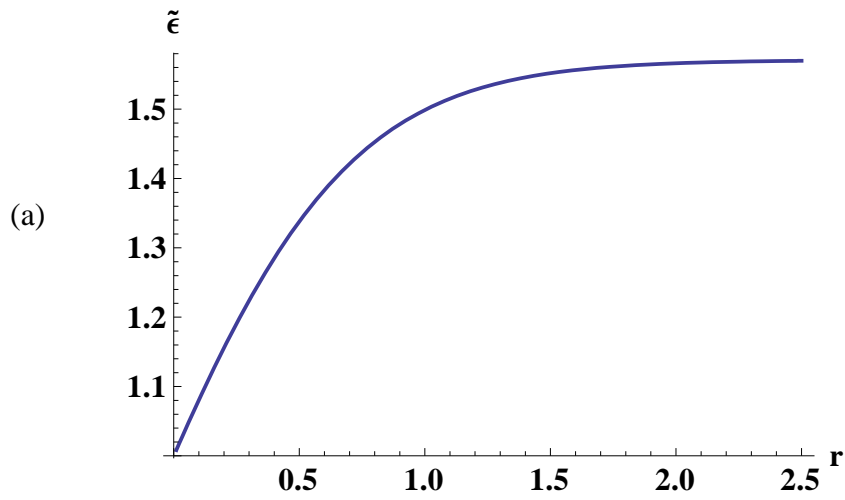

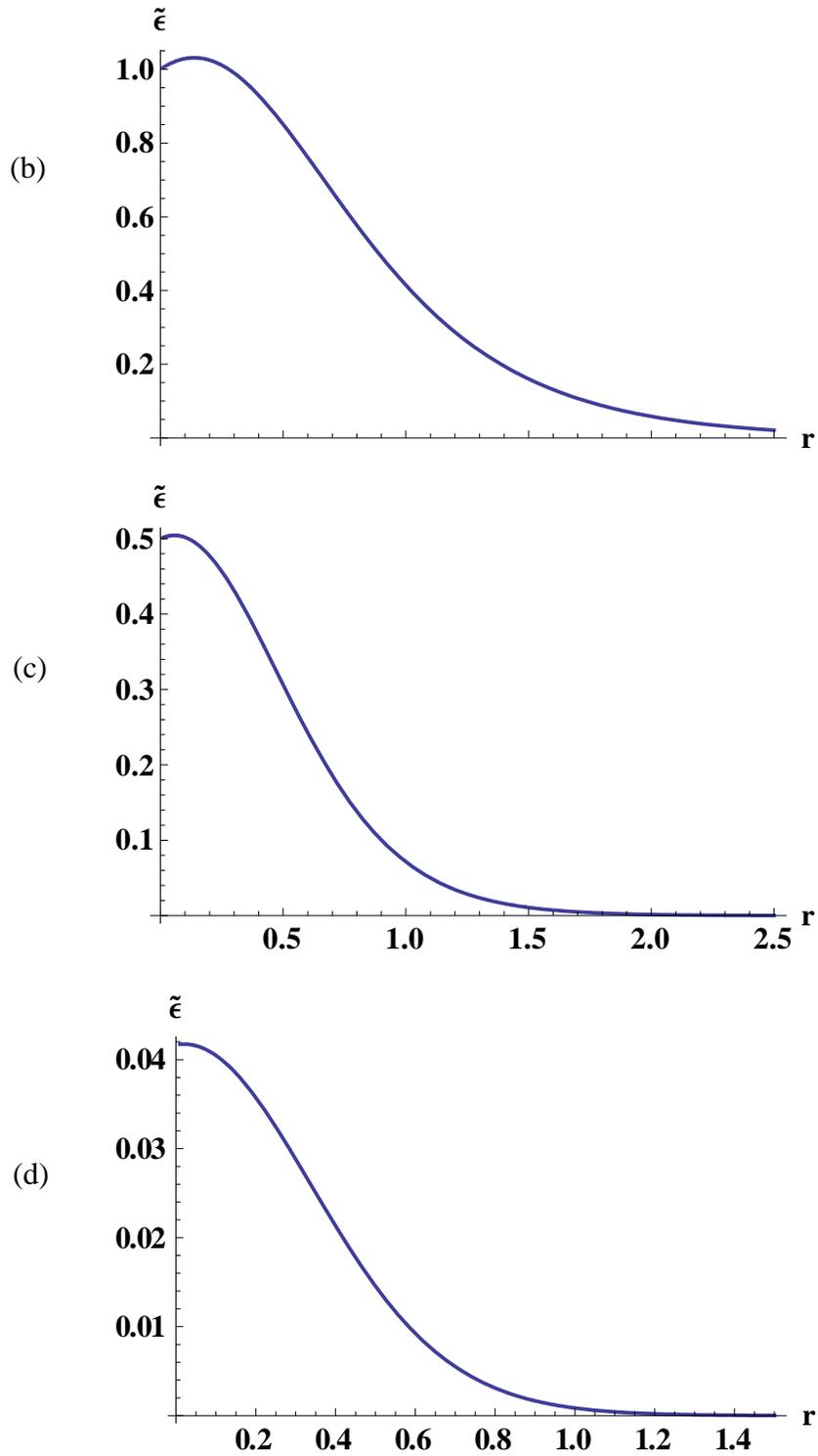

**Figure 2:** The ratio $\tilde{\varepsilon}$ as a function of *r* for photon subtractions from one of the modes of $\xi\rangle$ (a) single photon ( it reproduces Fig 3 of [9]), (b) two photon, (c) three photon, (d) four photon.